\documentclass[aps,prd,preprint,tightenlines,superscriptaddress,nofootinbib,dvips]{revtex4}
\usepackage{amsmath,amssymb}
\usepackage{graphicx,tabularx}

\renewcommand{\textfraction}{0.0}

\def\bq{\begin{equation}}
\def\eq{\end{equation}}
\def\ba{\begin{eqnarray}}
\def\ea{\end{eqnarray}}

\begin{document}

\preprint{$
\begin{array}{l}
\mbox{UB-HET-09-01}\\
\mbox{February~2009}\\[5.mm]
\end{array}
$}

\date{\today}

\title{Higgs Boson Exchange Effects in $l^+l^-\to W^+W^-$ at High Energy}

\author{U.~Baur\footnote{baur@ubhex.physics.buffalo.edu}}
\affiliation{Department of Physics,
State University of New York, Buffalo, NY 14260, USA}

\author{E.~Brewer\footnote{evbrewer@buffalo.edu}}
\affiliation{Department of Physics,
State University of New York, Buffalo, NY 14260, USA}

\begin{abstract}
\vspace{5mm}

We consider the prospects for detecting effects due to the Higgs
exchange diagram in high energy ${\mu}^+ {\mu}^-$, $e^+ e^-$, and
${\tau}^+ {\tau}^-$ collisions producing a pair of $W$ bosons.  The
processes $l^+l^- \rightarrow W^+W^-$ (with $l=\mu,e,\tau$) are
analyzed, analytically and via numerical simulations, to
determine the center of mass energy, $\sqrt{s}_H$, where the effects from Higgs
exchange become relevant.  The scaling of $\sqrt{s}_H$ with the mass of
the incoming 
leptons is also studied.  Special consideration is given to the $W^+W^-
\rightarrow l^{\pm} {\nu}_l jj$ final state after experimental
acceptance cuts are imposed.  Angular cuts are shown to be able to
significantly lower $\sqrt{s}_H$.
\end{abstract}

\maketitle

\newpage


\section{Introduction}
\label{sec:sec1}

Although the Standard
Model (SM)~\cite{Weinberg:1967tq,Salam:1968rm,Glashow:1961tr} of particle
physics is extremely successful in describing elementary particles
and their interactions (except gravity), one of the
key particles predicted and required by the SM to explain
the origin of mass, the so called Higgs boson, still remains elusive. 
The importance of the Higgs boson in the SM is not limited
to the gauge invariant generation of particle masses.
For processes like $e^+e^- \rightarrow W^+W^-$, the coupling of the Higgs
boson to fermions and $W$ bosons is necessary to maintain 
S-matrix unitarity. Unitarity of the S-matrix
reflects the requirement of probability conservation and requires that
partial wave amplitudes behave like $E^{\alpha}$ ($\alpha < 0$) at high
energies, $E$, for  
renormalizable theories~\cite{Llewellyn_Smith:1973ey,Joglekar:1973hh}.
Logarithmically growing terms are also allowed 
since they may be canceled by higher order 
corrections~\cite{Cornwall:1974co} and, thus, do not spoil
renormalizability. 

In 1974, Joglekar~\cite{Joglekar:1973hh} showed that S-matrix 
unitarity forces the couplings of the electroweak gauge bosons and the
Higgs boson to take the form of SM couplings at asymptotically high
energies.  This implies that, for non-zero lepton masses, Higgs boson
exchange has to contribute to the process $l^+l^-\to W^+W^-$ ($l=
e,\mu,\tau$), and that 
the coupling of the Higgs boson to leptons and $W$ bosons has to be of
SM form in order for S-matrix unitarity to be maintained.

In this paper we investigate through analytical calculations and
numerical simulations at what center of mass energy, $\sqrt{s}_H$, the
Higgs boson 
exchange diagram in  $l^+l^-\to W^+W^-$ becomes important. In
particular we investigate how experimental acceptance cuts on the $W$ decay
products affect $\sqrt{s}_H$. With linear
$e^+e^-$ colliders in the $0.5-3$~TeV energy range on the drawing
board~\cite{Brau:2007zza,Braun:2008zz}, and active development of a muon
collider with center of 
mass energies in the multi-TeV range ongoing~\cite{Ankenbrandt:2007zz},
the question 
whether one may be able to detect Higgs boson exchange effects in $l^+l^-\to
W^+W^-$ is of interest. 

The remainder of this paper is
organized as follows. In Sec.~\ref{sec:sec2} we discuss in some detail
the analytical calculation, and give a brief overview of how our
numerical simulations were performed. Results of the numerical
simulations are presented in Sec.~\ref{sec:sec3}. We concentrate on
$\mu^+\mu^-$ collisions, but also comment on the $e^+e^-$ case, and, for
completeness, on the academic case of $\tau^+\tau^-$ 
collisions\footnote{$\tau^+ \tau^-$ collisions are of academic interest
only, since the $\tau$ lepton is too short lived for efficient
acceleration and collimation into a beam.}. 
We summarize our results in Sec.~\ref{sec:sec4}.

\section{Details of the Numerical and Analytical Calculation}
\label{sec:sec2}

To determine the center of mass energy for which the Higgs exchange
diagram becomes important for maintaining S-matrix unitarity in
$l^+l^- \rightarrow W^+ W^-$, we calculate the cross section with and
without Higgs boson exchange. Including the Higgs boson exchange diagram
lowers the total cross section for $W$ pair production in lepton
collisions. To quantify for which center of mass energy the Higgs
exchange diagram becomes relevant, we impose simple requirements which
are discussed in more detail below. 

If $W$ decays are not taken into account, the calculation is
simple enough to be carried out analytically. The analytical
calculation is presented in Sec.~\ref{sec:sec2a}. However, for a more
realistic estimate, $W$ decays should be taken into account. If both
$W$'s decay leptonically, the final state contains two neutrinos which
both escape undetected. This complicates event
reconstruction. The small branching ratio of $WW\to
l\nu_ll'\nu_{l'}$ constitutes an additional disadvantage of the
all-leptonic final state. If both $W$'s decay hadronically, the QCD
$l^+l^-\to 4$~jet process represents a potentially worrisome
background. We therefore concentrate on the $WW\to l\nu jj$ final state
which has a large branching ratio, manageable background, and can be fully
reconstructed. To calculate the cross section for $l^+l^-\to W^+W^-\to
l\nu jj$, we use the parton level event generator {\tt
MadEvent}~\cite{Maltoni:2002qb}. Details of our {\tt MadEvent}
calculation are given in Sec.~\ref{sec:sec2b}.

\subsection{Analytical Calculation}
\label{sec:sec2a}

%
\begin{table}
\begin{center}
\setlength{\extrarowheight}{4pt}
\begin{tabular}{|c|}
\hline $d$-functions    \\
     \hline $d^2_{1,2}=-d^2_{-1,-2} = \frac{1}{2} (1 + \cos \theta) \sin
\theta$ \\ 
            $d^2_{1,-2}=-d^2_{-1,2} = -\frac{1}{2} (1 - \cos \theta)
\sin \theta$ \\ 
            $d^1_{1,1}=d^1_{-1,-1} = \frac{1}{2} (1 + \cos \theta)$ \\
            $d^1_{1,-1}=d^1_{-1,1} = \frac{1}{2} (1 - \cos \theta)$ \\
            $d^1_{1,0}=-d^1_{-1,0} = -\frac{1}{\sqrt{2}} \sin \theta$ \\
     \hline
\end{tabular}
\end{center}
\caption[]{$d$-functions used in the calculation of the helicity
amplitudes given by Eq.~(\ref{eq:amplitude})~\cite{Hagiwara:1986vm}.} 
\label{tab:dfunctions}
\vspace{5mm}
\end{table}

\begin{figure}[htbp]
\begin{center}
\includegraphics[viewport=50 120 612 700,width=15cm]{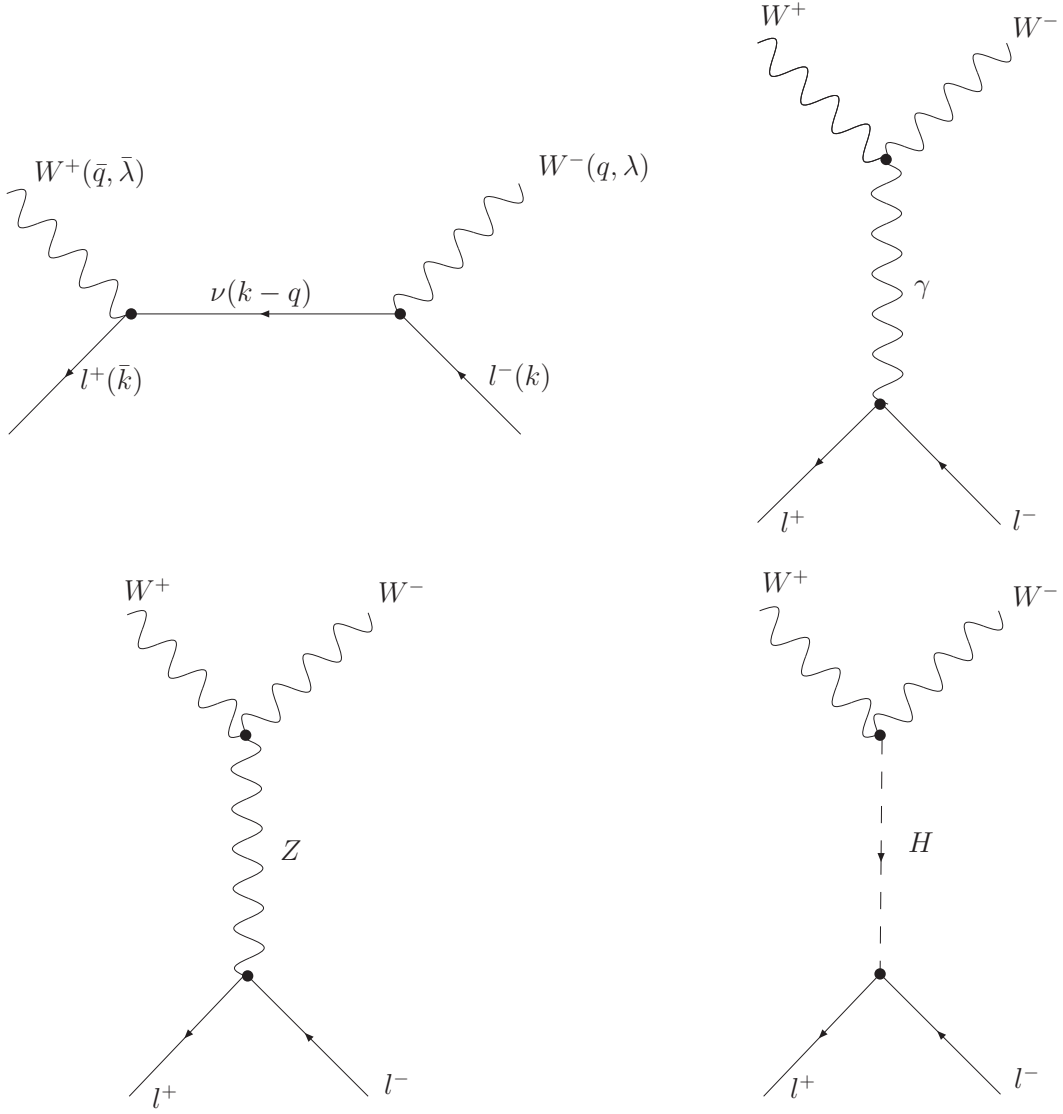}
\caption[]{The Feynman diagrams for $l^+l^- \rightarrow W^+W^-$.  } 
\label{fig:fig_llWW}
\vspace{-7mm}
\end{center}
\end{figure}

The Feynman diagrams for the process 
\begin{equation}\label{eq:llww-pol}
l^-(k, \sigma) + l^+(\bar{k},\bar{\sigma}) \rightarrow W^-(q,\lambda) +
W^+(\bar{q}, \bar{\lambda}) 
\end{equation}
are shown in Fig.~\ref{fig:fig_llWW}. 
Here, $k$ and $\bar{k}$ are the momenta of the incoming lepton and
anti-lepton, respectively, and $q$ and $\bar{q}$ are the momenta of the
$W^-$ and $W^+$ bosons. $\sigma$ and $\bar{\sigma}$ are the helicities of
the $l^-$ and $l^+$, and $\lambda$ and $\bar{\lambda}$ are the
polarizations of the $W$ bosons. The helicity amplitudes for $l^+l^-
\rightarrow W^+ W^-$ can be cast in the form~\cite{Hagiwara:1986vm}

\begin{equation}\label{eq:amplitude}
M_{\sigma \bar{\sigma};\lambda \bar{\lambda}}(\theta) = \sqrt{2} e^2
       \widetilde{M}_{\sigma \bar{\sigma};\lambda \bar{\lambda}}(\theta)
\Theta d^{J_0}_{\Delta \sigma, 
       \Delta \lambda} (\theta),
\end{equation}
where $e$ is the positron charge, 
$\Theta = \Delta \sigma (-1)^{\bar{\lambda}}$, $\Delta \sigma =
\frac{1}{2}(\sigma - \bar{\sigma})$, $\Delta \lambda = \lambda -
\bar{\lambda}$, $J_0 = \max(|\Delta \sigma|,|\Delta \lambda|)$ is the
minimum angular momentum of the system, and $\theta$ is the scattering
angle of the $W^-$ with respect to the $l^-$ direction in the $l^+ l^-$
center of mass frame.  $d^{J_0}_{\Delta\sigma, 
\Delta \lambda} (\theta)$ is the conventional $d$-function; its explicit
form for $J=1,2$ is given in Table~\ref{tab:dfunctions}. 
$\widetilde{M}_{\sigma \bar{\sigma};\lambda \bar{\lambda}}(\theta)$
denotes the remainder of the amplitude.

The explicit expressions for the helicity amplitudes in the high energy
limit obtained from the neutrino, photon and $Z$ exchange diagrams (see
Fig.~\ref{fig:fig_llWW}) are given by 
\begin{equation}\label{eq:M+-}
\begin{split}
M_{+-} =
\sqrt{2}e^2\frac{-\sqrt{2}}{\sin^2\theta_W}\frac{1}{2(1-\cos\theta)}
(-1)(-1)^{-1}d^2_{-1,2}\\
  = -\frac{e^2}{2\sin^2\theta_W}\sin\theta
\end{split}
\end{equation}
for transversely polarized $W$ bosons with ($(\lambda \bar{\lambda}) =
(+-)$) and $\Delta\lambda = +2$, 
\begin{equation}\label{eq:M-+}
\begin{split}
M_{-+} =
\sqrt{2}e^2\frac{-\sqrt{2}}{\sin^2\theta_W}\frac{1}{2(1-\cos\theta)}(-1)(-1)^{1}d^2_{-1,-2}\\
  = \frac{e^2}{2\sin^2\theta_W}\bigg( \frac{1+\cos\theta}{1-\cos\theta}\bigg)\sin\theta
\end{split}
\end{equation}
for transversely polarized $W$ bosons with ($(\lambda \bar{\lambda}) =
(-+)$) and $\Delta\lambda = -2$, and 
\begin{equation}\label{eq:M00tot}
M_{00} = -e^2 \bigg[ \frac{M^2_Z}{2M^2_W} -\frac{1}{2\sin^2\theta_W} \bigg(1 + \frac{M^2_Z}{2M^2_W} \bigg) \bigg] \sin\theta
\end{equation}
for longitudinally polarized $W$ bosons with ($(\lambda \bar{\lambda}) =
(00)$) and $\Delta\lambda = 0$. All other helicity amplitudes are
suppressed by a factor $\frac{1}{\gamma}$ or $\frac{1}{(\gamma)^2}$ with
$\gamma = E_W / M_W$ and, thus, can be ignored in the high energy limit.
Furthermore, the $\sin\theta$ kinematic factor softens the amplitudes given
by Eqs.~(\ref{eq:M+-}) and~(\ref{eq:M00tot}), making them numerically
smaller than the amplitude given in Eq.~(\ref{eq:M-+})
for a wide range of scattering angles. In Eqs.~(\ref{eq:M+-})
--~(\ref{eq:M00tot}), $M_W$ ($M_Z$) denotes the mass of the $W$ ($Z$)
boson, $e$ is the positron charge, and $\theta_W$ is the weak mixing angle.

When non-zero lepton masses are taken into account, the neutrino
diagram produces an extra term which grows proportional to 
$E$~\cite{Joglekar:1973hh} in the high energy limit. This term is
canceled by the Higgs exchange diagram. The Higgs exchange
amplitude is given by 
\begin{equation}\label{eq:higgscalc}
M_{higgs} = \frac{i}{\sqrt{2}}y^A_l u(k) \bar{v}(\bar k)
\frac{1}{(q+\bar q)^2-m^2_h+i\Gamma_hm_h}
            \frac{e}{\sin\theta_W}M_W g^{\alpha \alpha'}
\epsilon^{\lambda}_{\alpha}(q) 
	    \epsilon^{\lambda'}_{\alpha'}(\bar q),
\end{equation}
where $m_l = m_e, m_{\mu}, m_{\tau}$,
\begin{equation}\label{eq:yakawa}
y^A_l = \frac{e}{\sqrt{2}\sin\theta_W \cos\theta_W} \frac{m_l}{M_Z}
\end{equation}
is the Yukawa coupling, $u(k)$ and $\bar{v}(\bar k)$ are the lepton
spinors, $m_h$ is the mass of the Higgs boson, and
$\epsilon^{\lambda}_{\alpha}(q)$ and
$\epsilon^{\lambda'}_{\alpha'}(\bar q)$ are the polarization vectors of the
$W^-$ and $W^+$, respectively. $\Gamma_h$, finally, is the width of the
Higgs boson. For the Higgs boson masses currently favored by
experimental data~\cite{ewk08}, $\Gamma_h\ll m_h$. We, therefore, ignore
the width of the Higgs boson in the following discussion.

Squaring the amplitude in
Eq.~(\ref{eq:higgscalc}), averaging over the lepton spins and summing over
$W$ bosons polarizations results in the following expression 
\begin{equation}\label{eq:M-h-sq-mand}
\begin{split}
\frac{1}{4} \sum_{\lambda,\lambda', spins} |M_{higgs}|^2 = \frac{1}{2}
\bigg( \frac{y^A_l e M_W}{\sin\theta_W}\bigg)^2 \bigg(
\frac{\big(\frac{s}{2}-M^2_W \big)^2-M^4_W}{M^4_W} \bigg) 
\bigg( \frac{1}{s-m^2_h} \bigg)^2 \bigg( \frac{s}{2} - 2 m^2_l \bigg) ,
\end{split}
\end{equation}
where $s=(q+\bar q)^2$ is the squared center of mass energy. In the high
energy limit, 
the expression in Eq.~(\ref{eq:M-h-sq-mand}) simplifies to 
\begin{equation}\label{eq:M-h-sq-final}
\frac{1}{4} \sum_{\lambda,\lambda', spins} |M_{higgs}|^2 =
\bigg( \frac{y^A_l e }{4 M_W \sin\theta_W}\bigg)^2 s =
\bigg( \frac{e^4 m^2_l}{32 M^4_W \sin^4\theta_W}\bigg) s.
\end{equation}

One can use the expression given in Eq.~(\ref{eq:M-h-sq-final}) to
estimate $\sqrt{s}_H$.
The Higgs exchange diagram becomes important when $M_{higgs}$ and the
amplitude originating from the remaining three diagrams are of the same
order. Since 
$M_{-+}$ and $M_{+-}$ dominate over a wide range of scattering angles at high
energies, we can get a rough idea 
at what energies the Higgs exchange diagram becomes important 
by setting Eq.~(\ref{eq:M-h-sq-final}) equal to the sum of the squared 
helicity amplitudes $|M_{-+}|^2+|M_{+-}|^2$, averaged over the fermion spins,
(see Eq.~(\ref{eq:M-+})): 
\begin{equation}\label{eq:criteria}
\frac{1}{4} \sum_{\lambda,\lambda', spins} |M_{higgs}|^2 =
\frac{1}{4}\sum_{spins}\left(|M_{-+}|^2+|M_{+-}|^2\right ), 
\end{equation}
where
\begin{equation}\label{eq:M-+sq}
\frac{1}{4}\sum_{spins}\left( |M_{-+}|^2+|M_{+-}|^2\right ) =
\frac{e^4}{16\sin^4\theta_W}\left [\bigg( 
\frac{1+\cos\theta}{1-\cos\theta}\bigg)^2+1\right ]\sin^2\theta. 
\end{equation}

\begin{figure}[t!]
\begin{center}
\begin{tabular}{lr}
\includegraphics[width=8.2cm]{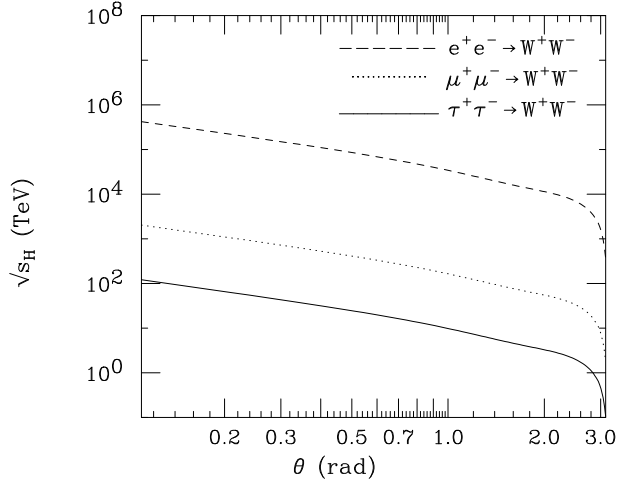}\\
\end{tabular}
\caption[]{Shown is the dependence of the center of mass energy
$\sqrt{s}_H$ for which Higgs exchange becomes important on the
scattering angle $\theta$.  Results are 
shown for $e^+e^- \rightarrow W^+W^-$, $\mu^+\mu^- \rightarrow W^+W^-$,
and $\tau^+\tau^- \rightarrow W^+W^-$.} 
\label{fig:CMEvsTHETA}
\vspace{-7mm}
\end{center}
\end{figure}

The center of mass energy for which the Higgs exchange diagram becomes
important is then given by 
\begin{equation}\label{eq:s}
\sqrt{s}_H = \frac{\sqrt{2} M^2_W}{m_l} \sin\theta\,\sqrt{\bigg(\frac{1 +
\cos\theta}{1-\cos\theta} \bigg)^2+1}~,
\end{equation}
{\it ie.} it scales like $1/m_l$.
Figure~\ref{fig:CMEvsTHETA} shows $\sqrt{s}_H$ as a function of the
scattering angle $\theta$. The
values for $\sqrt{s}_H$ obtained for $\theta=\pi / 2$ are listed in
Table~\ref{tab:s_90}. 

\renewcommand{\bottomfraction}{0.9}
\renewcommand{\textfraction}{0}

\begin{table}
\begin{center}
\setlength{\extrarowheight}{4pt}
\begin{tabular}{|c|c|}
\hline process & $\sqrt{s}_{H(\pi/2)}$ (TeV)\\
     \hline $\tau^+ \tau^- \rightarrow W^+ W^-$ & $7.3$ \\
            $\mu^+ \mu^- \rightarrow W^+ W^-$   & $120$ \\
	    $e^+ e^- \rightarrow W^+ W^-$       & $2.5 \times 10^4$ \\
     \hline
\end{tabular}
\end{center}
\caption[Center of mass energies for which $\frac{1}{4}
\sum_{\lambda,\lambda', spins} |M_{higgs}|^2 =$ \newline
$\frac{1}{4}\sum_{spins}|M_{-+}|^2+\frac{1}{4}\sum_{spins}|M_{+-}|^2$
for $\theta = \pi/2$.]{Center of mass energies for which $\frac{1}{4}
\sum_{\lambda,\lambda', spins} |M_{higgs}|^2
=\frac{1}{4}\sum_{spins}\left(|M_{-+}|^2+|M_{+-}|^2\right)$ 
for $\theta = \pi/2$.}
\label{tab:s_90}
\vspace{5mm}
\end{table}
%

\subsection{Simulation of $l^+l^- \rightarrow W^+W^-$ in MadEvent}
\label{sec:sec2b}

The results shown in Fig.~\ref{fig:CMEvsTHETA} and Table~\ref{tab:s_90}
indicate that Higgs boson exchange becomes relevant at energies much
higher than those foreseen for future $e^+e^-$ and $\mu^+\mu^-$
colliders. However, these results do not take
into account interference effects between the Higgs exchange diagram and
the other three diagrams. Furthermore, $W$ decays, and effects caused by
experimental cuts, are not included.

These effects can easily be taken into account in numerical
simulations. We have used the tree-level event generator
{\tt MadEvent}~\cite{Maltoni:2002qb} to perform simulations of the
process $l^+l^-\to W^+W^-\to l'\nu_{l'}jj$. {\tt MadEvent} assumes
electrons and muons to be massless. Since lepton masses are essential
when taking into account the Higgs exchange diagram in $l^+l^-\to
W^+W^-$, we modified the {\tt MadEvent} source code to include finite
masses for electrons and muons. We only consider the $l'\nu_{l'}jj$
final state. As discussed in Sec.~\ref{sec:sec1}, the $l'\nu_{l'}jj$
final state can easily be reconstructed, has a fairly large branching
ratio, and a relatively small background.

The SM parameters and cuts we used for our simulations are
given in Table~\ref{tab:parameters}.  The Higgs boson mass of
$m_h=129$~GeV was arbitrarily chosen from the accepted Higgs mass range
($114~{\rm GeV}<m_h<185$~GeV at $95\%$ $CL$)~\cite{ewk08}.
As a quantitative measure for estimating $\sqrt{s}_H$, we require that
the cross section 
with and without Higgs boson exchange differ by a factor~2: 
\begin{equation}\label{eq:crossectioncriteria100}
{\sigma}_{\rm (without~Higgs)} = 2{\sigma}_{\rm(with~Higgs)}.
\end{equation}

\begin{table}
\begin{center}
\setlength{\extrarowheight}{4pt}
\begin{tabular}{|c|c|}
\hline  parameter    &  value \\
     \hline $G_F$  & $1.16637$  \\
     \hline $\alpha(M_Z)$ & $1/128.9$ \\
     \hline $\sin^2 \theta_W$ & $0.23153$ \\
     \hline $M_W$ & $80.425$~GeV \\
     \hline $M_Z$ & $91.1876$~GeV \\
     \hline $m_t$ & $172.7$~GeV   \\
     \hline $m_b$ & $4.3$~GeV \\
     \hline $m_c$ & $1.2$~GeV \\
     \hline $m_{\tau}$ & $1.7769$~GeV \\
     \hline $m_e$ & $0.511$~MeV \\
     \hline $m_{\mu}$ & $0.10566$~GeV\\
     \hline $m_{h}$ & $129$~GeV \\
     \hline $\tau$ width & $2.36 \times 10^{-12}$~GeV \\
     \hline
\end{tabular}
\end{center}
\caption{SM parameters used for {\tt MadEvent} simulations.}
\label{tab:parameters}
\vspace{5mm}
\end{table}

To simulate detector response, we impose acceptance cuts on the final
state particles. For definiteness, we chose cuts similar to those imposed
by the LEP experiments~\cite{Bardin:1997gc}. We shall comment below how
our results change if these cuts are modified. 

It should be noted that the criteria for estimating $\sqrt{s}_H$ given
in Eqs.~(\ref{eq:criteria}) and~(\ref{eq:crossectioncriteria100}) are
not identical. In addition to the dominant helicity amplitudes $M_{+-}$
and $M_{-+}$, Eq.~(\ref{eq:crossectioncriteria100}) includes the
contributions of $M_{00}$, $M_{++}$ and $M_{--}$, as well as interference
effects between the Higgs exchange amplitude and the other amplitudes. It
will be interesting to compare the numerical results obtained using
Eq.~(\ref{eq:criteria}) and~(\ref{eq:crossectioncriteria100}).

\section{Numerical Results}
\label{sec:sec3}

%
\renewcommand{\bottomfraction}{0.9}
\renewcommand{\textfraction}{0}
\begin{figure}[t!]
\begin{center}
\begin{tabular}{lr}
\includegraphics[width=10.0cm]{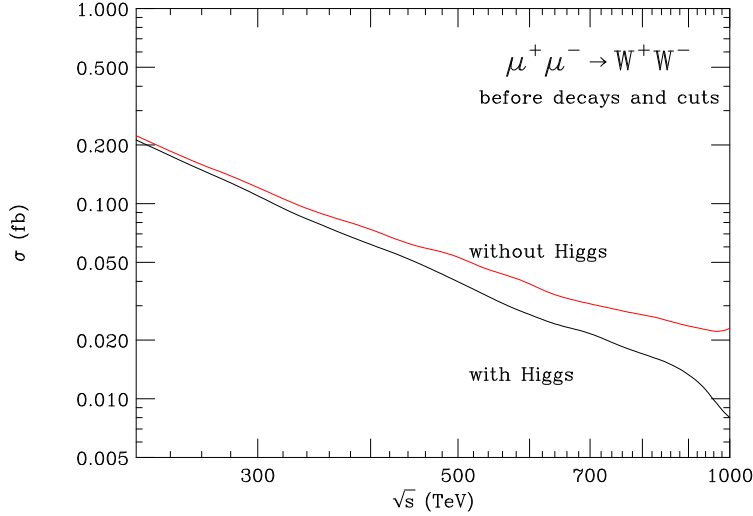}\\
\\
\end{tabular}
\caption[The total cross section for ${\mu}^+{\mu}^- \rightarrow W^+W^-$
before decays and cuts]{The total cross section for the
process ${\mu}^+{\mu}^- \rightarrow W^+W^-$ calculated with (black line)
and without (red line) the Higgs exchange diagram included as a function
of the center of mass energy, $\sqrt{s}$. } 
\label{fig:mmWWnodecays}
\vspace{-7mm}
\end{center}
\end{figure}
\renewcommand{\bottomfraction}{0.9}
\renewcommand{\textfraction}{0}
\begin{figure}[t!]
\begin{center}
\begin{tabular}{lr}
\includegraphics[width=10.0cm]{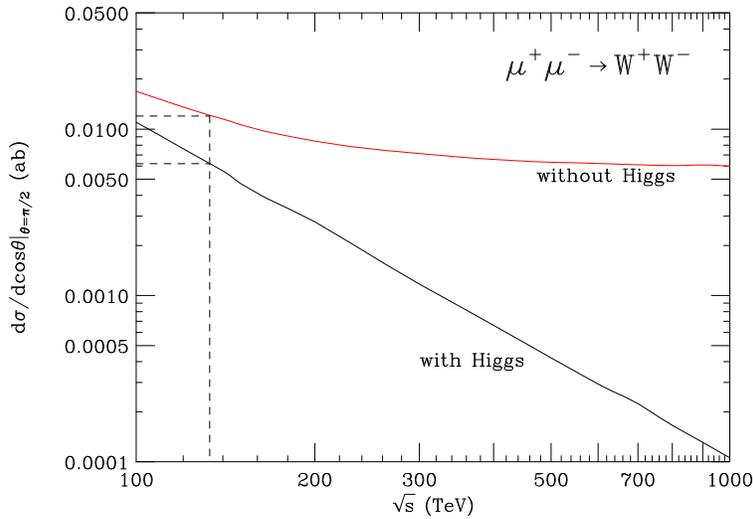}\\
\\
\end{tabular}
\caption[The total cross section for ${\mu}^+{\mu}^- \rightarrow W^+W^-$
with pseudorapidity cuts on $W$ bosons]{The differential cross
section $d\sigma/d\cos\theta$ for ${\mu}^+{\mu}^- \rightarrow W^+W^-$
with $\theta=\pi/2$ calculated 
with (black line) and without (red line) the Higgs exchange diagram
included.  The dashed lines indicate
the center of mass energy ($\sqrt{s}=130$~TeV) where the cross
sections differ by a factor~2.} 
\label{fig:mmWWnodecays90}
\vspace{-7mm}
\end{center}
\end{figure}

We illustrate the effect of the Higgs exchange diagram on the $l^+l^-\to
W^+W^-$ total cross section in Fig.~\ref{fig:mmWWnodecays} for the
$\mu^+\mu^-$ case. As expected from the analytic estimate (see
Fig.~\ref{fig:CMEvsTHETA}), the Higgs exchange diagram becomes important
in the few hundred TeV region; Eq.~(\ref{eq:crossectioncriteria100}) is
satisfied for $\sqrt{s}\approx 900$~TeV. Similar calculations performed
for $e^+e^-$ and $\tau^+\tau^-$ collisions also confirm the results of
Fig.~\ref{fig:CMEvsTHETA}, in particular the $1/m_l$ scaling of
$\sqrt{s}_H$. 

The analytic estimate, Eq.~(\ref{eq:s}), gives the center of mass energy
as a function of the scattering angle, determined directly from a
comparison of the squared amplitudes. Since the squared amplitude is
proportional to the differential cross section $d\sigma/d\cos\theta$, it
is useful to impose Eq.~(\ref{eq:crossectioncriteria100}) on the
differential cross section, and then compare with the result obtained
from Eq.~(\ref{eq:s}). Figure~\ref{fig:mmWWnodecays90} shows
$d\sigma/d\cos\theta$ for $\theta=\pi/2$ as a function of the center of
mass energy with and without Higgs exchange. The dashed lines indicate
where the differential cross sections differ by a factor~2. 

The analytic estimate for $\theta=\pi/2$ is compared with the result
of the {\tt MadEvent} simulation for $e^+e^-$, $\mu^+\mu^-$ and
$\tau^+\tau^-$ collisions in Table~\ref{tab:results1}. Although the
criteria for obtaining $\sqrt{s}_{H(\pi/2)}^{analytic}$ and
$\sqrt{s}_{H(\pi/2)}^{MadEvent}$ 
are somewhat different, the numerical results agree at the 10\% level,
indicating that 
the sub-dominant amplitudes and interference effects play a minor role
only. In particular,
Table~\ref{tab:results1} confirms that the center of mass energy for
which Higgs boson exchange becomes relevant scales with $1/m_l$. 
\begin{table}
\begin{center}
\setlength{\extrarowheight}{4pt}
\begin{tabular}{|c|c|c|c|}
\hline  process    & $\sqrt{s}_{H(\pi/2)}^{analytic}$ (TeV) &
 $\sqrt{s}_{H(\pi/2)}^{MadEvent}$ (TeV) & percentage difference \\
     \hline ${\tau}^+ {\tau}^- \rightarrow W^+ W^-$  & $7.3$ & $8.0$ &
$9.6\%$ \\ 
     \hline ${\mu}^+ {\mu}^- \rightarrow W^+ W^-$ & $120$ & $130$ & $9.2\%$ \\
     \hline ${e}^+ {e}^- \rightarrow W^+ W^-$ & $2.5 \times 10^4$ & $2.8
\times 10^4$ & $10.4\%$ \\ 
     \hline
\end{tabular}
\end{center}
\caption{Comparison of the center of mass energy for which the Higgs
exchange diagram becomes relevant obtained analytically and from {\tt
MadEvent} simulations for $\theta=\pi/2$. } 
\label{tab:results1}
\vspace{5mm}
\end{table}

Comparing Figs.~\ref{fig:mmWWnodecays} and~\ref{fig:mmWWnodecays90} it
is obvious that the Higgs exchange diagram has a more pronounced effect
on the differential cross section at larger 
scattering angles than on the total cross section, which is dominated by
the contribution from small values of $\theta$. This is due to the fact
that the Higgs exchange diagram leads to an isotropic distribution of
the $W$ bosons and their decay products, whereas the contribution from
the remaining diagrams is strongly peaked at small scattering angles due
to the (massless) neutrino exchange $t$- and $u$-channel diagrams. It
also suggests that
experimental cuts, in particular angular cuts, could substantially
lower the center of mass energy for which Higgs boson exchange becomes
important in $l^+l^-\to W^+W^-\to l'\nu_{l'}jj$. To be specific, we impose the
following cuts~\cite{Bardin:1997gc} in our subsequent discussion:
\begin{equation}
E_l>1~{\rm GeV}, \qquad E_{jet}> 3~GeV.
\end{equation}
Furthermore, the scattering angle of the leptons has to be in the
range 
\begin{equation}
10^\circ<\theta_l<170^\circ,
\end{equation}
and the opening angle between a lepton and a jet has
to be $\theta_{l,jet}>5^\circ$. Finally, the invariant mass of the two
jets has to be $m_{jj}>5$~GeV. 

\begin{figure}[t!]
\begin{center}
\begin{tabular}{lr}
\includegraphics[width=13.0cm]{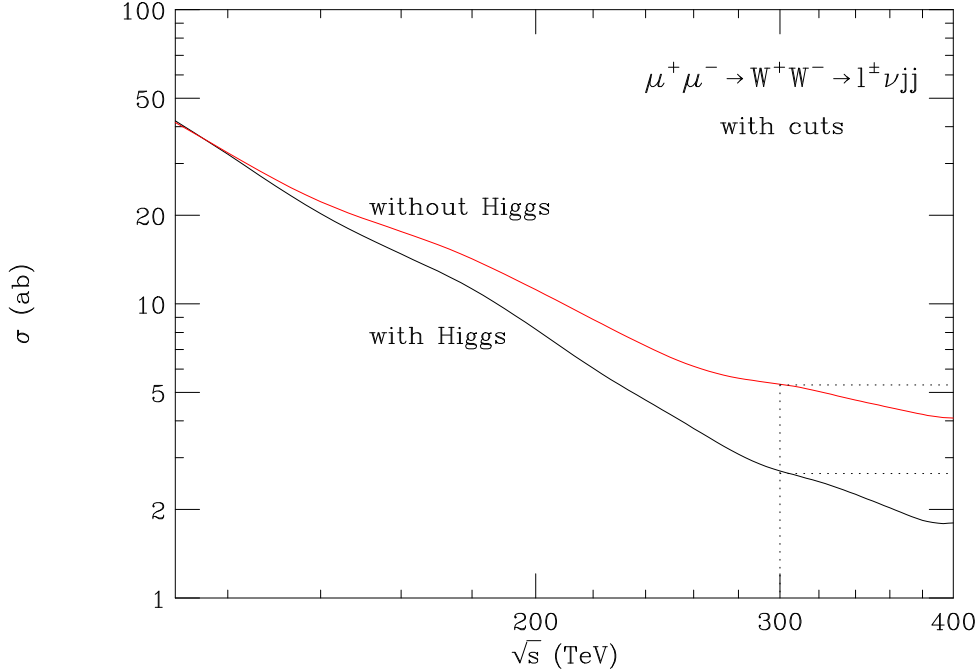}\\
\\
\end{tabular}
\caption[The total cross section for ${\mu}^+{\mu}^- \rightarrow W^+W^-
\rightarrow l^{\pm} {\nu}_l jj$ before experimental cuts ($100\%$
difference)]{The cross section for the process
${\mu}^+{\mu}^- \rightarrow W^+W^- \rightarrow l^{\pm} {\nu}_l jj$
calculated with the Higgs (solid line) and without the Higgs (red
line) exchange diagram included.  The dotted lines indicate
the center of mass energy ($\sqrt{s}=300$~TeV) where the cross
sections differ by a factor~2. The cuts imposed on the final state
particles are discussed in the text.} 
\label{fig:mmWWwithcuts100}
\vspace{-7mm}
\end{center}
\end{figure}

Figure~\ref{fig:mmWWwithcuts100} shows the ${\mu}^+{\mu}^- \rightarrow
W^+W^- \rightarrow l^{\pm} {\nu}_l jj$ cross section, including angular
and energy cuts on the final state particles, as a function of
the center of mass energy with and without Higgs boson exchange. The center
of mass energy for which ${\sigma}_{\rm{(without~Higgs)}} =
2{\sigma}_{\rm{(with~Higgs)}}$ now is $\sqrt{s}_H\approx 300$~TeV, about a
factor~3 less than without cuts. 
\renewcommand{\bottomfraction}{0.9}
\renewcommand{\textfraction}{0}

For $e^+e^-$ ($\tau^+\tau^-$) collisions, $\sqrt{s}_H$ is larger
(smaller) than for $\mu^+\mu^-$ collisions. Since the $l^+l^-\to W^+W^-$
cross section becomes increasingly peaked at smaller $W$ scattering
angles, and thus 
at smaller lepton and jet scattering angles, one expects that the
angular cuts will lower $\sqrt{s}_H$ by a larger (smaller) amount for $e^+e^-$
($\tau^+\tau^-$) collisions. Explicit calculations show that this is
indeed the case. Table~\ref{tab:results2} compares the values of
$\sqrt{s}_H$ obtained with and without cuts for $e^+e^-$, $\mu^+\mu^-$,
and $\tau^+\tau^-$ collisions. 

\begin{table}
\begin{center}
\setlength{\extrarowheight}{4pt}
\begin{tabular}{|c|c|c|c|}
\hline collisions    &  $\sqrt{s}_{H}^{MadEvent}$ (before
decays and cuts) & $\sqrt{s}_{H}^{MadEvent}$ (after decays and
cuts) & ratio
 \\ 
     \hline ${\tau}^+ {\tau}$  & $50$~TeV           & $20$~TeV
& $2.50$ \\ 
     \hline ${\mu}^+ {\mu}^-$  & $940$~TeV         & $300$~TeV
& $3.13$ \\ 
     \hline ${e}^+ {e}^-$      & $2.0 \times 10^5$~TeV & $3.0 \times
10^4$~TeV & $6.67$ \\ 
     \hline
\end{tabular}
\end{center}
\caption{Comparison of the center of mass energy for which
${\sigma}_{\rm (without~Higgs)} = 2{\sigma}_{\rm(with~Higgs)}$  
before and after $W$ decays and acceptance cuts are taken into account.} 
\label{tab:results2}
\vspace{5mm}
\end{table}

\begin{figure}[t!]
\begin{center}
\begin{tabular}{lr}
\includegraphics[width=7.0cm]{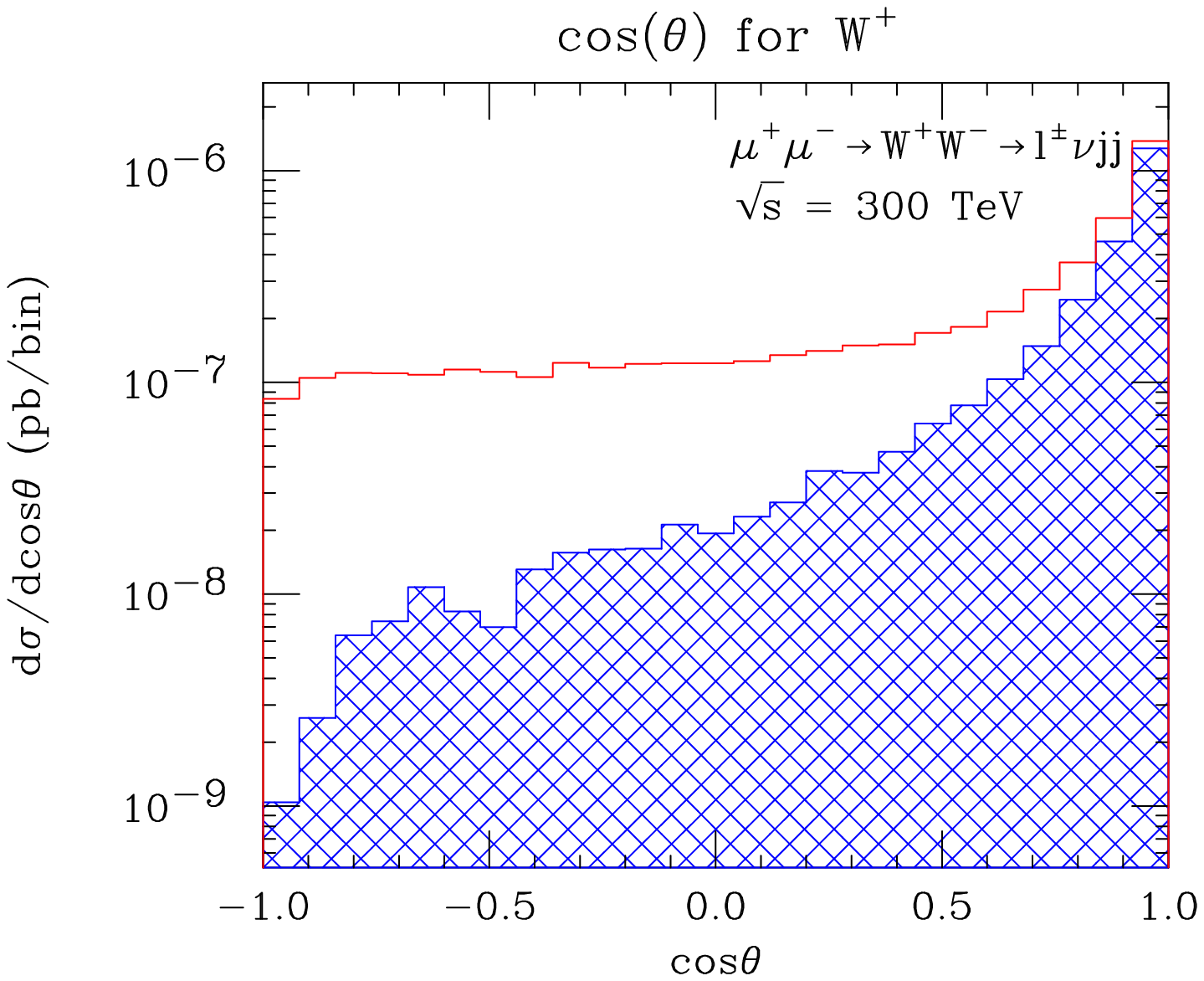} &
\includegraphics[width=7.0cm]{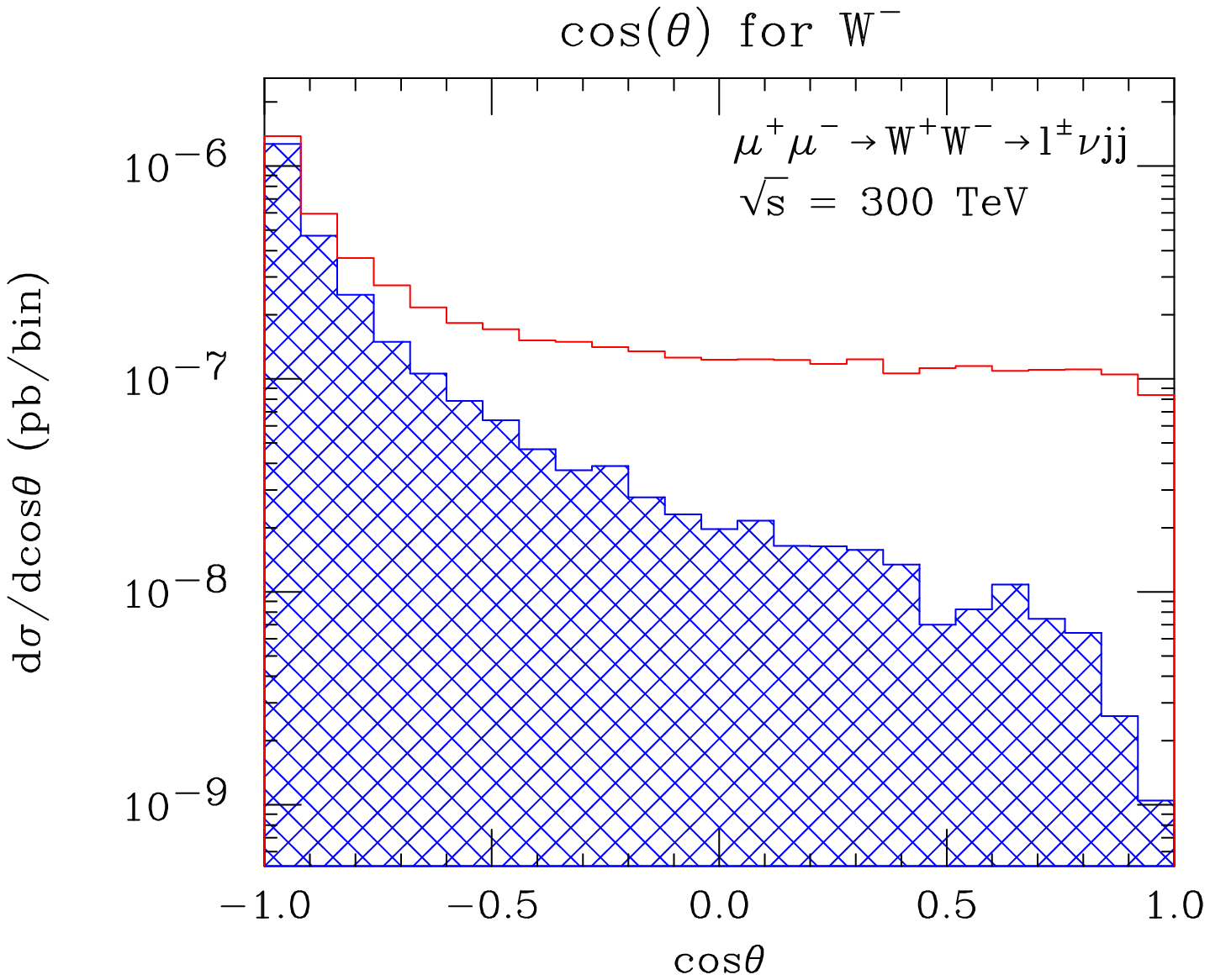} \\
(a) & (b) \\
\\
\end{tabular}
\caption[]{The differential cross section for the $W^+$ and
$W^-$ bosons in ${\mu}^+{\mu}^- \rightarrow W^+W^- \rightarrow l^{\pm}
{\nu}_l jj$ for $\sqrt{s} = 300$~TeV calculated with (blue hatched
histogram) and without (red histogram) the Higgs exchange diagram,
averaged over 100 simulations each containing several thousand of 
events~\cite{Brewer:2008ev}.} 
\label{fig:mmWWcos1}
\vspace{-7mm}
\end{center}
\end{figure}

\begin{figure}
\begin{center}
\begin{tabular}{lr}
\includegraphics[width=7.0cm]{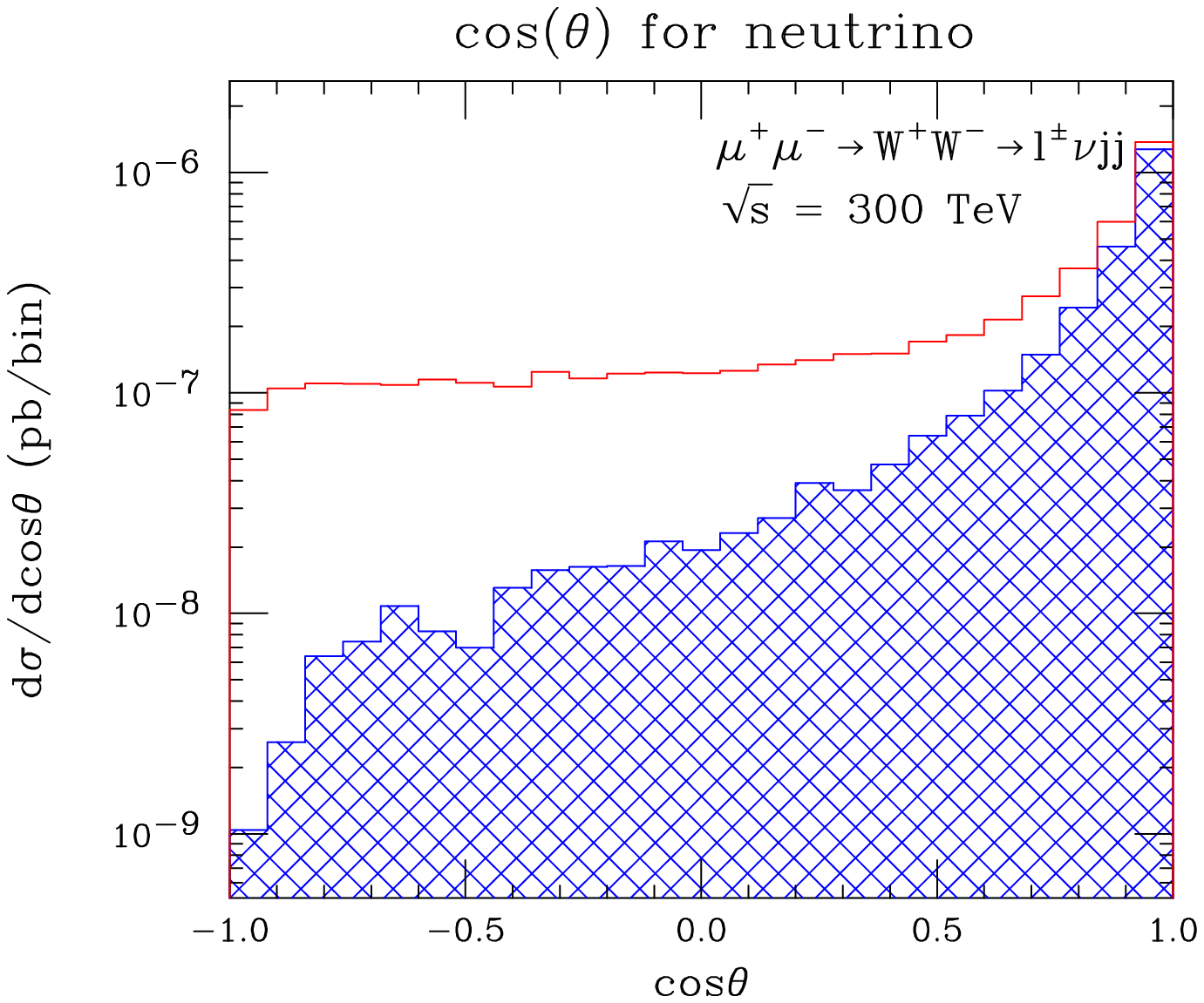} &
\includegraphics[width=7.0cm]{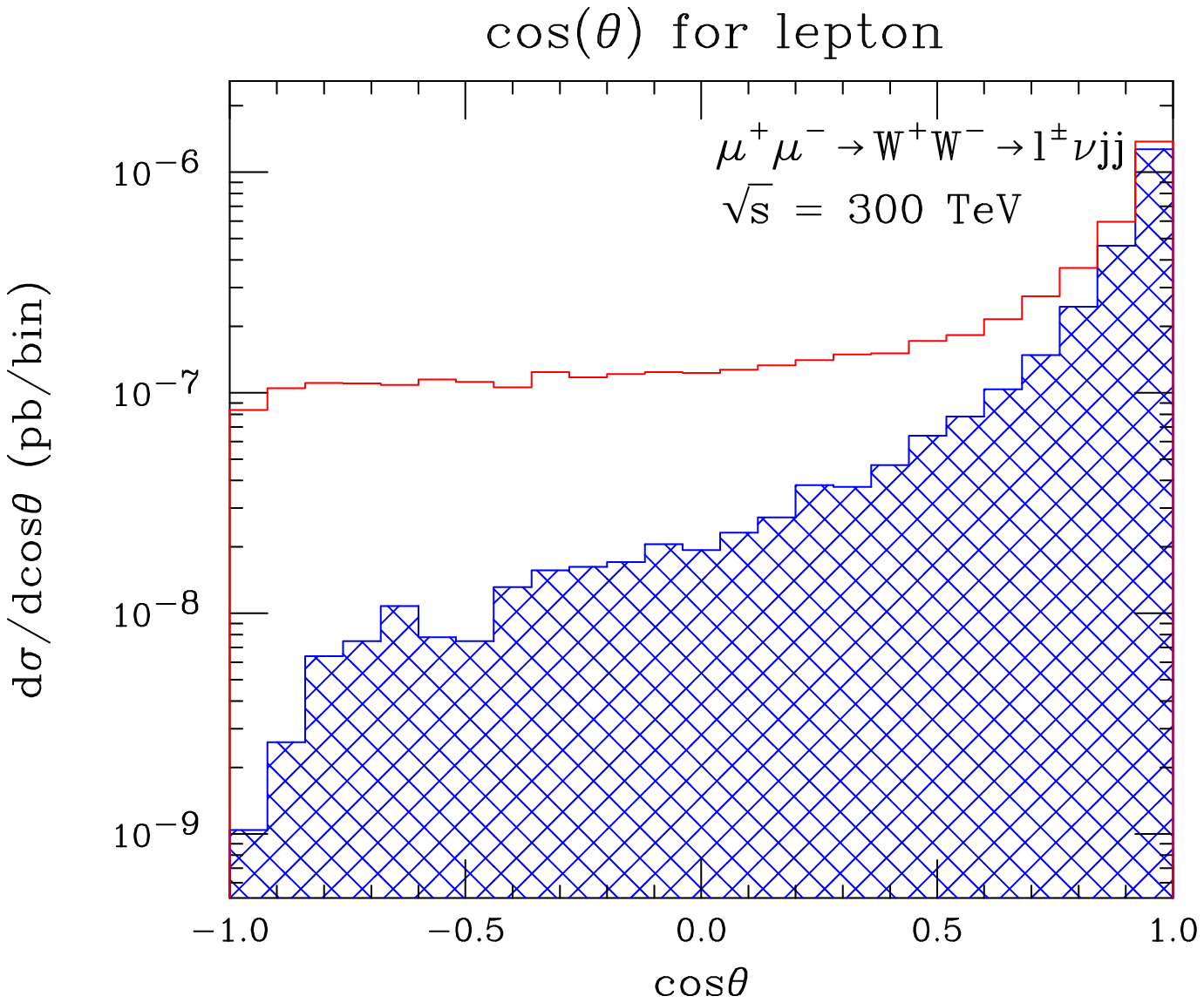} \\
(a) & (b) \\
\includegraphics[width=7.0cm]{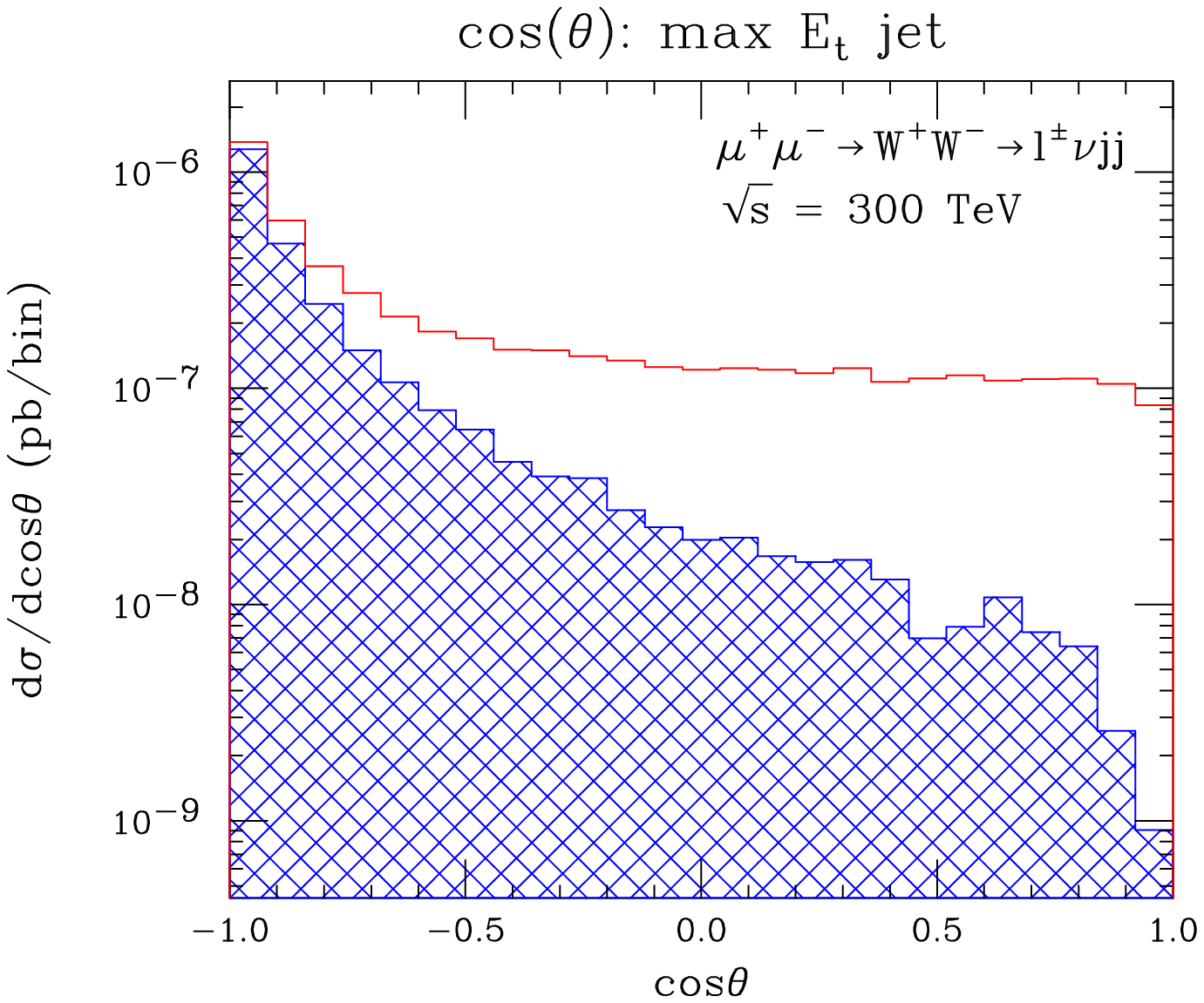} &
\includegraphics[width=7.0cm]{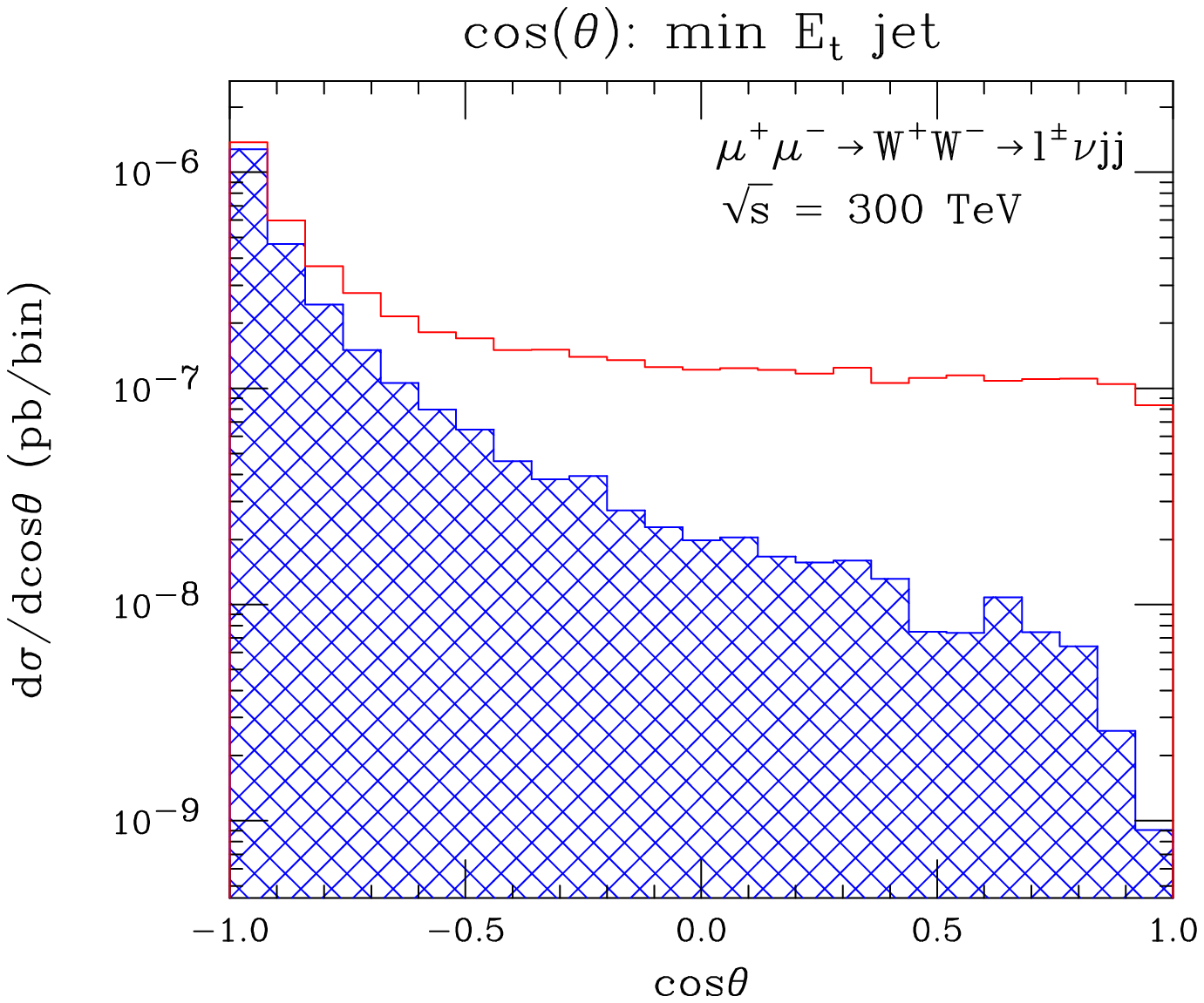} \\
(c) & (d) \\
\\
\end{tabular}
\caption[]{The differential cross sections for the final state
particles in ${\mu}^+{\mu}^- \rightarrow W^+W^- \rightarrow l^{\pm}
{\nu}_l jj$ for $\sqrt{s} = 300$~TeV, calculated with (blue
hatched histogram) and without (red histogram) the Higgs exchange
diagram, averaged over 100 simulations each containing thousands of 
events~\cite{Brewer:2008ev}.} 
\label{fig:mmWWcos2}
\vspace{-7mm}
\end{center}
\end{figure}

The effect of the Higgs exchange diagram can be seen in more detail in
the angular distributions of the $W$ bosons, and their decay products
which are shown for ${\mu}^+{\mu}^- \rightarrow W^+W^- \rightarrow l^{\pm}
{\nu}_l jj$ and $\sqrt{s} = 300$~TeV in Figs.~\ref{fig:mmWWcos1}
and~\ref{fig:mmWWcos2}. 
\renewcommand{\bottomfraction}{0.9}
\renewcommand{\textfraction}{0}
\renewcommand{\bottomfraction}{0.9}
\renewcommand{\textfraction}{0}
The Higgs exchange diagram is seen to have only a small effect at small
scattering angles, but becomes much more important for larger values of
$\theta$. For $\cos\theta\approx 1$ ($\cos\theta\approx -1$) of the
$W^-$ ($W^+)$ scattering angle, the Higgs exchange 
diagram reduces the differential cross section by almost two
orders of magnitude. Because of the V-A nature of the
$Wl\nu$ coupling, the $W$ decay products inherit the characteristics of
the parental differential cross section, i.e. the differential cross
sections for the final state neutrino (Fig.~\ref{fig:mmWWcos2}(a)) and
corresponding lepton (Fig.~\ref{fig:mmWWcos2}(b)) are similar to that
of the $W^+$ (see Fig.~\ref{fig:mmWWcos1}(a)).  Most of the final state
leptons and neutrinos scatter close to the beam direction.  Likewise,
for the $W^-$ decaying into two jets, the angular distributions of the
jet with maximum transverse energy (Fig.~\ref{fig:mmWWcos2}(c)) and jet
with minimum transverse energy (Fig.~\ref{fig:mmWWcos2}(d)) are similar
to the angular distribution of the $W^-$ (Fig.~\ref{fig:mmWWcos1}(b)).
Because of this similarity between the distributions of the $W$ bosons
and the final state particles, a relatively larger percentage of the
final state particles will scatter close to the beam direction when the
Higgs diagram is taken into account than when it is
omitted~\cite{Brewer:2008ev}. Thus, imposing experimental acceptance
cuts will reduce the differential cross section with the Higgs diagram
taken into account by a larger percentage than without. 

\begin{figure}[t!]
\begin{center}
\begin{tabular}{lr}
\includegraphics[width=7.0cm]{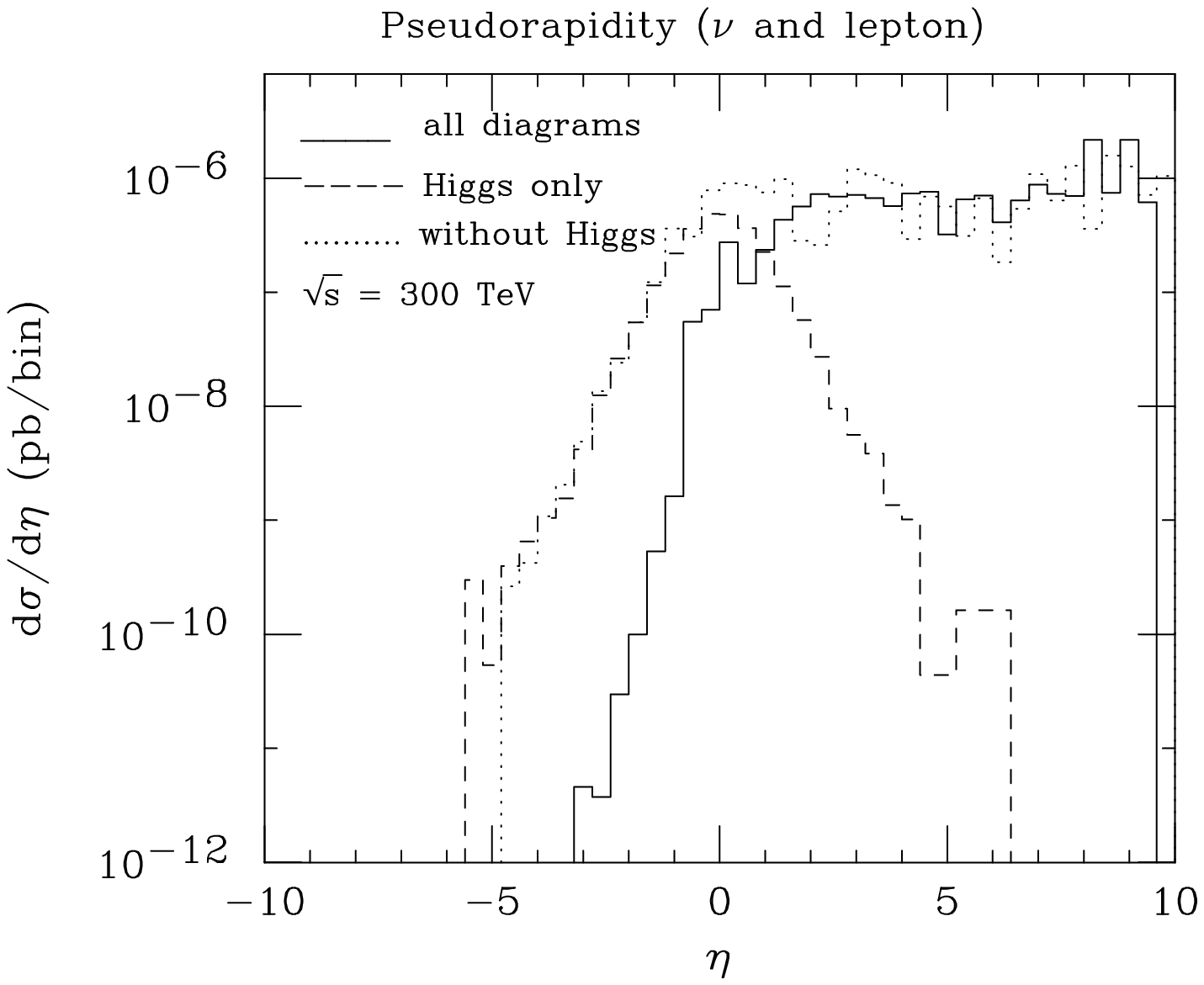} &
\includegraphics[width=7.0cm]{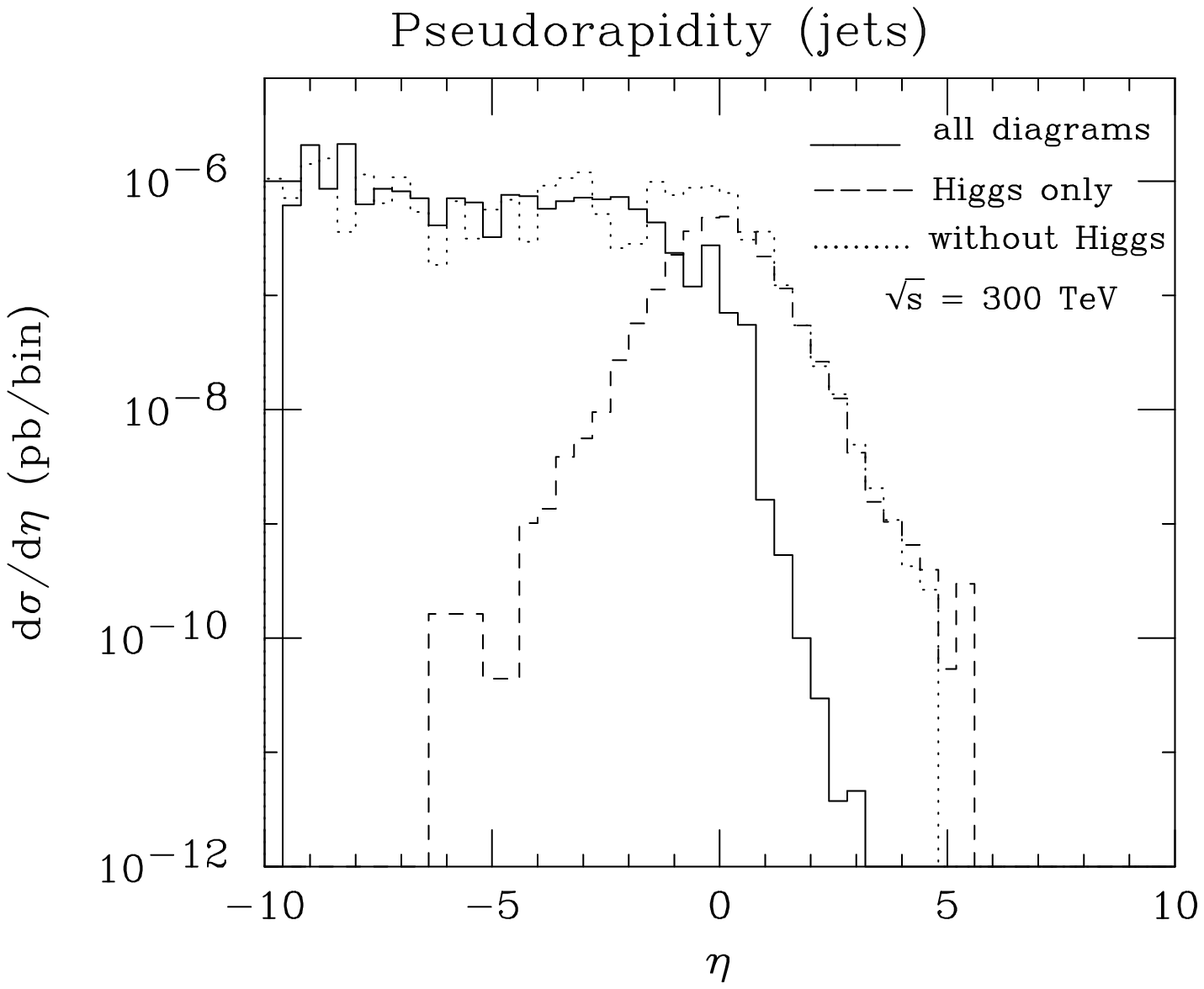} \\
(a) & (b) \\
\\
\end{tabular}
\caption[]{Differential cross sections as a function of the
pseudorapidity of the final state particles for ${\mu}^+{\mu}^-
\rightarrow W^+W^- \rightarrow l^{\pm} {\nu}_l jj$ before cuts. Results
are shown for all diagrams (solid line) in Fig.~\ref{fig:fig_llWW}, for
diagrams (a-c) in Fig.~\ref{fig:fig_llWW} (dotted line), and for diagram
(d) from Fig.~\ref{fig:fig_llWW} (dashed line). All calculations are
performed for $\sqrt{s} = 300$~TeV~\cite{Brewer:2008ev}.} 
\label{fig:mmWWrap2}
\vspace{-7mm}
\end{center}
\end{figure}

A more detailed view is offered in
Fig.~\ref{fig:mmWWrap2}, where the pseudorapidity, $\eta$, of the final
state particles is shown for ${\mu}^+{\mu}^-
\rightarrow W^+W^- \rightarrow l^{\pm} {\nu}_l jj$ and
$\sqrt{s}=300$~TeV without imposing any cuts. Here, the pseudorapidity
is defined by
\begin{equation}
\eta=-\log\left[\tan\left(\frac{\theta}{2}\right)\right].
\end{equation}
\renewcommand{\bottomfraction}{0.9}
\renewcommand{\textfraction}{0}

Based on the angular distributions for ${\mu}^+{\mu}^- \rightarrow
W^+W^- \rightarrow l {\nu}_l jj$, we can draw the following
conclusions~\cite{Brewer:2008ev}: 
\begin{itemize}
   \item {The angular cuts imposed on the final state particles remove
events where the final state particles scatter close to the beam
($\theta \le 10^{\circ}$ or ($\theta \ge 170^{\circ}$).  This means that
a substantial portion of the 
total cross section is discarded because of the strong forward peaking of the
cross section.} 
   \item {The discarded events constitute a higher percentage of
${\sigma}_{\text{with Higgs}}$ than of ${\sigma}_{\text{without Higgs}}$
due to the fact that most of the cancellations between $M_{Higgs}$ and
$M=M_{\nu}+M_{\gamma}+M_Z$ occur at large scattering angles. This is a
result of the spin~0 nature of the
Higgs boson: $s$-channel Higgs exchange leads to an isotropic
distribution for the $W$ 
bosons. Due to the $V-A$ character of the $Wl\nu$ coupling, the final
state leptons and jets largely inherit the characteristics of the
angular distribution of their $W$ parents. } 
   \item{ Thus, experimental cuts will increase the difference
between the ${\sigma}_{\text{with Higgs}}$ and ${\sigma}_{\text{without
Higgs}}$ for 
any particular center of mass energy and shift the values of $\sqrt{s}_H$.} 

\end {itemize}

\renewcommand{\bottomfraction}{0.9}
\renewcommand{\textfraction}{0}
\begin{figure}[t!]
\begin{center}
\begin{tabular}{lr}
\includegraphics[width=13.0cm]{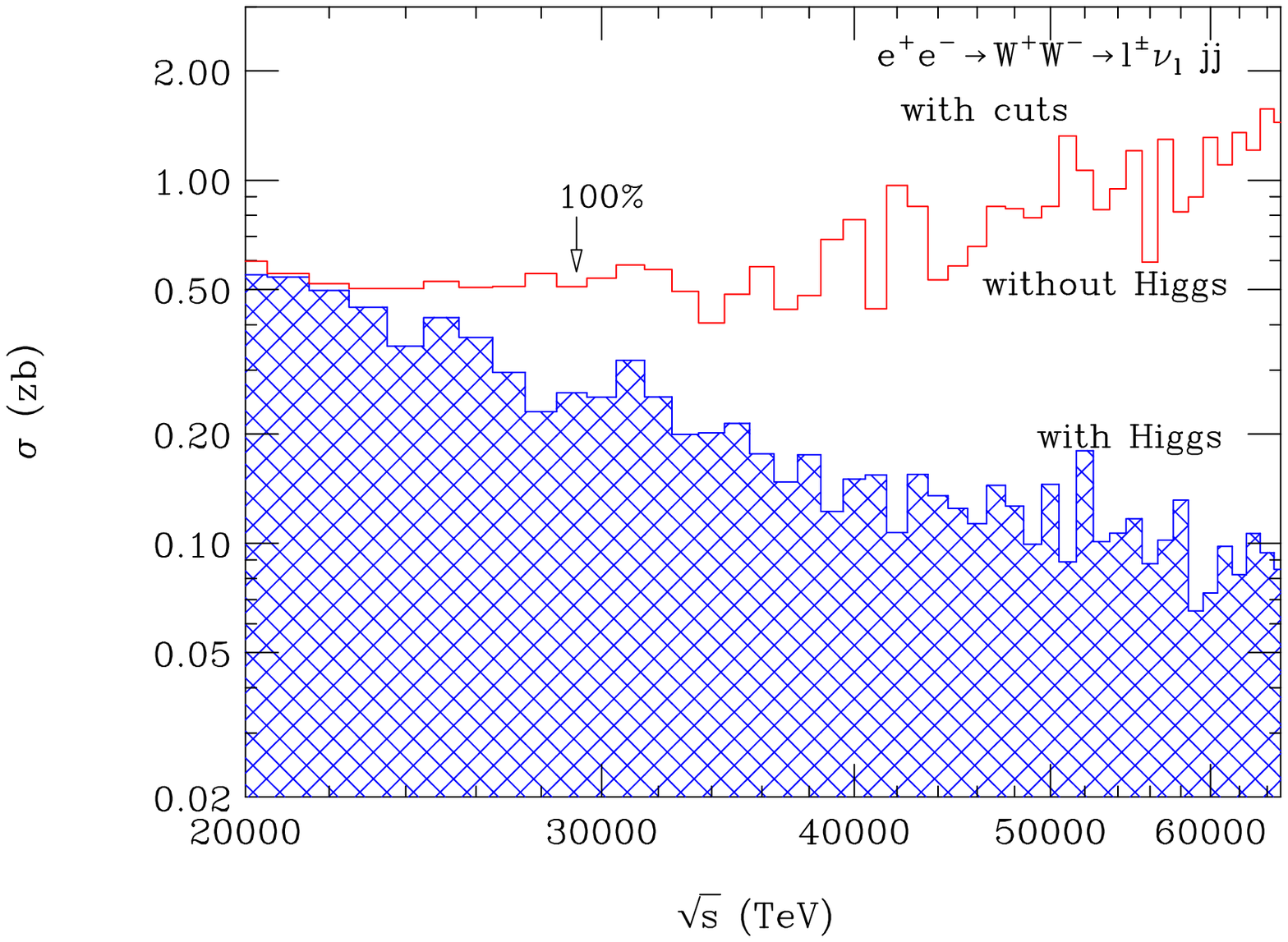}\\
\\
\end{tabular}
\caption[]{Shown are the total cross sections for the process
${e}^+{e}^- \rightarrow W^+W^- \rightarrow l^{\pm} {\nu}_l jj$
calculated with the Higgs (hatched blue histogram) and without the Higgs
(red histogram) exchange diagram.  The arrow indicates the center of
mass energy where the cross sections differ by $100\%$ ($\sqrt{s}\approx
3.0\times 10^4$~TeV).
For $\sqrt{s}\ge 3.6 \times 10^4$~TeV the total cross section
calculated without the Higgs exchange diagram begins to rise and,
eventually, will violate the bound imposed by S-matrix
unitarity~\cite{Brewer:2008ev}.}  
\label{fig:eeWWwithcuts}
\vspace{-7mm}
\end{center}
\end{figure}

Qualitatively similar results are obtained for $e^+e^-$ and
$\tau^+\tau^-$ collisions. The results obtained for the $e^+e^-\to
W^+W^-\to l\nu jj$ cross section with cuts imposed deserve special
attention. 
Fig.~\ref{fig:eeWWwithcuts} shows that the total cross section for the
process calculated without the Higgs exchange diagram begins to increase
for $\sqrt{s} \ge 3.6 \times 10^4$~TeV, which explicitly indicates
that S-matrix unitarity is violated. 

We now briefly comment on the uncertainties in the simulated data and in
the cross section obtained by MadEvent for ${\mu}^+{\mu}^- \rightarrow
W^+W^- \rightarrow l^{\pm} \nu_l jj$.  Large cancellations between the
diagrams result in large uncertainties for both ${\sigma}_{\text{with
Higgs}}$ and ${\sigma}_{\text{without Higgs}}$.  At $\sqrt{s}=300$~TeV
the standard deviation for ${\sigma}_{\text{with Higgs}}=5.4$~ab is
$0.38$~ab which corresponds to about $7\%$ of the total cross 
section.  For the same center of mass energy, the standard deviation for
${\sigma}_{\text{without Higgs}}=2.7$~ab is $0.28$~ab corresponding
to $10.4\%$ of the total cross section.  In both cases, the
uncertainties in the cross section are large, due to cancellations
between the individually divergent amplitudes $M=M_{\nu}+M_{\gamma}+M_Z$
and $M_{Higgs}$.  The uncertainties for electron and $\tau$-lepton
collisions are similar. 

Finally, we briefly comment on how higher order radiative corrections
may affect our results. The full NLO electroweak corrections to
$l^+ l^- \rightarrow W^+ W^- \rightarrow l'^{\pm} \nu_{l'} jj$ including
lepton mass effects (and the Higgs exchange diagram) have not been calculated
yet.  The complexity of the NLO electroweak corrections does not allow
us to make educated guesses about the possible effect of these
corrections on $\sqrt{s}_H$. However, at high energies, higher order
electroweak corrections are known to
grow as $\log \frac{s}{M^2_W}$~\cite{Kuroda:1990wn}, and eventually have
to be resummed. This may considerably 
change the numerical results presented here. The energy 
for which Higgs exchange becomes relevant may well increase or decrease by
a factor of~2 or more once these effects are taken into account.


\section{Summary and Conclusions}
\label{sec:sec4}


In this paper, we have considered the high energy limit of the processes
$l^+l^- \rightarrow W^+ W^- \rightarrow l' 
{\nu}_{l'} jj$ ($l = e, {\mu}, {\tau}$) in
order to determine the center of mass energy, $\sqrt{s}_H$, for which
the Higgs exchange diagram becomes relevant. In the SM, the Higgs
exchange diagram is needed in order to guarantee that S-matrix unitarity
is maintained in the high energy limit.
Two, slightly different, criteria were used to estimate $\sqrt{s}_H$,
both leading to similar results. 

From a simplified analytical calculation we found that 
$\sqrt{s}_H$ is of order $M_W^2/m_l$, where $m_l$ is the mass of the
incoming lepton. As a result, the center of mass energies for which
Higgs boson exchange becomes important for $e^+e^-$, $\mu^+\mu^-$ and
$\tau^+\tau^-$ collisions are in the region of ${\cal O}(10
- 10^5$~TeV), which is far above the energy range of any $\mu^+\mu^-$ or
$e^+e^-$ collider currently on the drawing board. 

We also investigated how experimental acceptance cuts influence the
center of mass energy for which Higgs boson exchange becomes
important. The $t$- and $u$-channel fermion exchange diagrams result in
a strong peaking of the $l^+l^-\to W^+W^-$ cross section at small
scattering angles, 
which, due to the $V-A$ nature of the $W$-fermion coupling, is inherited
by the $W$ decay products. The Higgs exchange diagram, on the other
hand, leads to a distribution which peaks at large scattering
angles. Imposing angular cuts on the final state particles thus tends to
lower the center of mass energy for which the Higgs exchange diagram
becomes important. The effect of the angular cuts increases with
growing energies, and 
is most pronounced for $e^+e^-$ collisions where the angular cuts chosen
in this paper decrease
$\sqrt{s}_H$ by almost a factor~7. For comparison, for $\mu^+\mu^-$
($\tau^+\tau^-$) collisions, angular cuts decrease $\sqrt{s}_H$ by about
a factor~3 (2.5). For more (less) stringent angular cuts, the effect on
$\sqrt{s}_H$ increases (decreases).

Our results were obtained from simple tree level calculations and one
does have to worry about how higher order electroweak corrections may
affect them. Unfortunately, the NLO electroweak corrections to
$l^+l^-\to W^+W^-\to 4$~fermions have been computed in the limit of
massless incoming leptons only~\cite{Denner:1999dt}. However,
electroweak radiative corrections are known to increase logarithmically
with the center of mass energy~\cite{Kuroda:1990wn}, and, eventually,
have to be 
resummed. They may thus substantially change $\sqrt{s}_H$, although the
general order of magnitude estimate presented in this paper should
remain correct. 

\acknowledgements
We would like to thank C.~Quigg for suggesting the problem. One of us
would like to thank the 
Fermilab Theory Group and the Physics Department, Michigan State
University, where part of this work was done, for 
their generous hospitality. This research was supported by the
National Science Foundation under grants No.~PHY-0456681 and PHY-0757691.



\end{document}